\documentclass[
manuscript,
amsmath,amssymb,aps]{revtex4-2}

\usepackage{graphicx}
\usepackage{dcolumn}
\usepackage{bm}
\usepackage[utf8]{inputenc}

\begin{document}

\title{Students' attitudes toward experimental physics in a conceptual inquiry-based introductory physics lab}

\author{Danny Doucette}
 \email{danny.doucette@pitt.edu, he/him/his}
\author{Russell Clark}
\author{Chandralekha Singh}
\affiliation{Department of Physics and Astronomy, University of Pittsburgh, Pittsburgh, Pennsylvania 15260, USA}

\date{\today}

\begin{abstract}

There is some evidence that conceptual inquiry-based introductory physics lab curricula, such as RealTime Physics, may improve students' understanding of physics concepts. Thus, these curricula may be attractive for instructors who seek to transform their physics labs to improve student learning. However, the impact of conceptual inquiry-based lab instruction on students' attitudes and beliefs about experimental physics, as measured by the E-CLASS survey, is not yet fully understood. We present data from three curricular approaches over four semesters ($n=701$). We saw no change in E-CLASS scores in the first implementation of a conceptual inquiry-based introductory physics lab. However, the addition of questions that asked students to reflect on issues relating to experimental physics was associated with E-CLASS outcomes that are comparable to other effective approaches to lab instruction.

\end{abstract}

\maketitle

\section{Introduction}

In recent years, there has been renewed interest in introductory college physics lab courses~\cite{HolmesLewandowskiLandscape}. This has led to the development of a variety of formats and pedagogical approaches for introductory physics labs~\cite{ISLE,SobhanzadehLabatorials} that mirrors the development and analysis of approaches for lecture-based classes~\cite{DocktorSynthesis,KalmanPeerInstruction,KarimEBAE}. In part, the interest in transforming lab instruction has been driven by research that suggests students do not learn physics concepts in traditional, highly-structured physics labs~\cite{HolmesValueAdded, SmithTeachingExperimentation}. In addition to addressing the need for students to learn physics concepts, another important course goal that motivates some lab transformations is the inclusion of scientific inquiry in lab experiences~\cite{AAPTlab,GalvezSinghThemeIssueIntro}. By inquiry, we refer to experimental work designed to probe and illuminate the validity and functioning of scientific models or hypotheses~\cite{AronsInquiry,KoponenGenerative}.

Corresponding to the various goals of lab transformations, a variety of evaluation tools have been developed and employed. Concept inventories such as the Force Concept Inventory (FCI)~\cite{HestenesFCI}, the Force and Motion conceptual Evaluation (FMCE)~\cite{ThorntonFMCE}, and the Mechanics baseline Test (MBT)~\cite{HestenesMBT} have been used to determine how much labs have helped students learn physics concepts. Assessments like Physics Lab Inventory of Critical Thinking (PLIC)~\cite{WalshHolmesPLIC} and Physics Measurement Questionnaire (PMQ)~\cite{VolkwynLabUncertainty} have been used to measure students' lab skills, such as critical thinking and uncertainty analysis. Surveys such as the Colorado Learning Attitudes About Science Survey for Experimental Physics (E-CLASS)~\cite{WilcoxECLASSValidation,ECLASSsummary} and the Maryland Physics Expectations Survey (MPEX)~\cite{RedishMPEX} have been used to evaluate changes in students' expectations about physics and attitudes toward aspects of experimental physics. It is noteworthy that traditional physics labs see a decrease in students' scores on the E-CLASS~\cite{WilcoxECLASSValidation} and other attitudinal surveys, suggesting that un-transformed lab instruction might push students to adopt less-expert-like views and attitudes about experimental science. 

One approach to transforming lab courses is to integrate lab and lecture, creating a learning environment that allows students to engage in scientific inquiry while also building up their understanding of physics concepts. Since inquiry is integrated into the course and classrooms are rearranged to promote collaborative work, such approaches typically do not have separate lab courses. Some examples of this approach include the Investigative Science Learning Environment (ISLE)~\cite{ISLE}, Modeling Instruction~\cite{BreweModeling}, Physics by Inquiry~\cite{PhysicsByInquiry}, Student-Centered Active Learning Environment with Upside-down Pedagogies (SCALE-UP)~\cite{BeichnerSCALEUP}, Studio Physics~\cite{LawsStudioPhysics}, or Technology-Enabled Active Learning (TEAL)~\cite{BelcherTEAL}. This approach has been associated with improved conceptual learning for students (e.g.,~\cite{EtkinaDesignReflection,DoriTEAL}), improved student attitudes toward science and learning science (e.g.,~\cite{BreweModelingCLASS,LindseyPhysByInqCLASS}), and unchanged (i.e., not decreasing) attitudes toward experimental physics~\cite{WilcoxImpactApproach}. Another approach, integrating lab and tutorial sections~\cite{AhrensmeirLabatorials,SobhanzadehLabatorials}, has likewise been associated with improved student outcomes~\cite{KalmanLabatorial}. However, due to financial, scheduling, or other constraints, it may not be possible for all institutions to switch from a lab-and-lecture introductory physics sequence to an integrated course. Thus, there is a need to examine student learning and attitudes toward science and experimental physics in stand-alone physics lab courses.

Little work has been done to evaluate how conceptual inquiry-based labs may affect students' attitudes toward experimental science. Student attitudes toward, and understanding of, experimentation can be an important learning outcome for lab courses, and can also moderate the effectiveness of lab work on conceptual learning. One study reported that E-CLASS scores increased in an inquiry-based lab but, as the authors argue, the context of that study makes it difficult to extrapolate~\cite{ShiGuidedInquiryECLASS}. A large-scale study by Wilcox and Lewandowski suggests that students' attitudes toward experimental science decrease in ``guided'' introductory physics labs and do not decrease in ``open-ended'' labs~\cite{WilcoxOpenEnded}. Another study using the same data shows that students' attitudes decrease in introductory physics labs in which the purpose is to ``reinforce physics concepts'' but increase very slightly in labs in which the purpose is to ``develop lab skills''~\cite{WilcoxDevelopingSkills}. However, since many conceptual inquiry-based labs do not fall neatly into the guided/open-ended/concepts/skills categorization scheme, it may be difficult to infer definite conclusions from this work about the impact of conceptual inquiry-based lab curricula on students' attitudes toward experimental physics. 

A secondary question relates to the impact of conceptual inquiry-based lab courses on student understanding of physics concepts. Several (non-integrated) transformed lab courses have adopted conceptual inquiry-based curricula as an alternative to a traditional, highly-structured lab curriculum. Conceptual inquiry-based lab curricula focus on providing students opportunities to conduct experiments as a way to develop their understanding of physics concepts~\cite{GalvezSinghThemeIssueIntro}. Evaluations of conceptual inquiry-based labs have reported on student conceptual understanding using the FCI~\cite{RiazDLSCL,NockInquiryLab,ChambersInquiryLab,VanDomelenLabFCI}, MBT~\cite{RiazDLSCL,NockInquiryLab}, and FMCE~\cite{RealTimePhysics}. In some studies showing a strong and positive impact of conceptual inquiry-based lab curricula on student outcomes, (e.g.,~\cite{RealTimePhysics}), student learning in lab and lecture are not separated out, making it difficult to infer the impact of a stand-along conceptual inquiry lab course. Generally, research focused just on learning in stand-alone labs suggests that conceptual inquiry-based curricula are responsible for a boost to student conceptual understanding, with effect sizes ranging from negligible to moderately positive. For example, Nock found that students who took an inquiry-based lab course outperformed students who took a traditional, highly-structured lab course on lecture test grades (81\% to 76\%), but that they performed equally well on the FCI~\cite{NockInquiryLab}. Likewise, Riaz et al.~\cite{RiazDLSCL} found no difference on FCI scores, but saw a significant difference on MBT scores for students in inquiry-based labs compared with students in traditional lab. Thus, it remains unclear whether conceptual inquiry-based labs may be a counterexample to recent scholarship~\cite{HolmesValueAdded,HolmesWiemanLabsNoEffect} that suggests traditional highly-structured lab courses are ineffective at helping students learn physics concepts. Research is needed to determine the magnitude of the difference between conceptual inquiry-based labs and traditional, highly-structured labs. Thus, while the main aim of this paper is to document how three different curricula may have influenced student attitudes toward experimental science, we also report the impact these curricula had on student understanding of physics concepts.

In this paper, we seek to address the research question: How are students' attitudes toward experimental physics impacted by a conceptual inquiry-based curriculum, as measured by the E-CLASS?. In addition, we seek to shed some light on two additional questions: What is the impact on students' attitudes of adding questions to a conceptual inquiry-based curriculum that ask students to reflect on the nature of their experimental work? And what is the impact on students' conceptual understanding in this conceptual inquiry-based introductory physics lab of different lab curricula? 

\section{Materials and Methods}

At many institutions, the introductory physics sequence is offered as a two-semester series of lectures with associated labs. In contrast, at our institution, a large public research-intensive university in the USA, the introductory physics lab is a single course that is not associated with the introductory lecture courses, although there is overlap in the physics concepts covered. The lab requires that students are simultaneously enrolled in, or have completed, the second semester of the introductory physics lecture course. Thus, all of the students who are enrolled in the lab have already completed the first half of the introductory physics sequence, and have completed their studies of the topic of mechanics.

In our labs, students collaborated in groups or 2 (or 3, if necessary), to conduct their experimental work. They wrote a group lab report during the three-hour lab period. Understanding the nature of student collaboration in introductory labs is beyond the scope of our analysis in this paper, but has been reported elsewhere~\cite{DoucetteHermione,Doucette4WomenEJP}.

\subsection{Three Lab Curricula}

In our study, students completed three different lab curricula. The first of these, which we will refer to as the traditional curriculum, is a set of highly-structured lab exercises that were written by the second author~\cite{ClarkOldlabs}. In the traditional curriculum, students followed step-by-step instructions as they set up apparatus, made measurements, and performed calculations. Measurements and calculations were submitted digitally using LON-CAPA~\cite{KortemeyerLONCAPA}, which performed some checks to ensure measured values were reasonable and calculated values were correct. Since they were given step-by-step instructions, students were able to operate sophisticated equipment, including oscilloscopes (measuring the speed of sound in a cardboard tube), teltron tubes (determining $q/m$ for an electron), and spectrometers (spectra from discharge tubes). The lab manual provided a review of the relevant physics each week, but students often found themselves ritualistically following the instructions from the lab manual rather than thinking about physics concepts. The traditional curriculum may be akin to introductory physics lab instruction as it has been practiced in many colleges and universities over the past century~\cite{BlessCookbook}, and is similar to the traditional, highly-structured lab courses analyzed by others (e.g.,~\cite{HolmesValueAdded,WilcoxImpactApproach}.

The second curriculum represents our best effort to faithfully implement 12 inquiry-based labs from the RealTime Physics curriculum~\cite{RealTimeBook1,RealTimeBook2,RealTimeBook3,RealTimeBook4}. The design of these labs was informed by physics education research and, indeed, RealTime Physics has been shown to boost student understanding of physics concepts~\cite{RealTimePhysics}. In our implementation of RealTime Physics labs, students completed a pre-lab activity individually, conducted the lab-work in pairs, and then completed a post-lab homework assignment individually. Each of these three tasks was graded. The lab-work called for students to develop multiple representations of physics concepts (e.g., ticker tape-style dot diagrams, velocity-time graphs, and written descriptions of motion). The lab-work frequently used a cycle whereby students made a prediction about the outcome of a simple phenomenon (e.g., what are the forces when two carts collide?), and then conducted the phenomenon to check their prediction (e.g., collide two carts with force probes on a dynamics track). The RealTime Physics labs are structured, in the sense that students complete a series of tasks, including stating predictions, conducting small experiments, and answering questions that connect experimental results to physics concepts. Students used computer-based data collection with Vernier equipment, including dynamics tracks, sensors, and optics benches. In our implementation, students completed 6 mechanics labs, 3 electric circuits and electromagnetism labs, and 3 optics labs, outlined in Table~\ref{labstable}. 

The third curriculum represents our efforts to adapt RealTime Physics to the needs of our students and our course goals, which were to improve student conceptual understanding of physics and to help students learn skills and ways of thinking relevant to experimental physics. Since we decided that the AAPT lab recommendations~\cite{AAPTlab} were well-aligned to our lab goals, we looked through the recommendations and wrote reflection questions that asked students to think about topics from the lab recommendations that were not otherwise addressed in our lab curriculum. For this third curriculum, we kept the pre-lab and post-lab assignments from RealTime Physics. We changed the lab-work that students performed in only one significant way, by adding reflection questions~\cite{KalmanInterventionsEpistemology} to the lab-work. These questions called for students to reflect on the nature of science and ways of thinking in experimental science, and are inspired by the American Association of Physics Teachers Recommendations for the Undergraduate Physics Laboratory Curriculum~\cite{AAPTlab} and recent scholarship about reflection and critical and scientific thinking in physics labs~\cite{DounasFrazerGuidedReflection,KoenigLabs,HolmesPNAS,RiosPathwaysLabs,Hu2017ECLASS}. These reflection questions asked students to think individually, confer with their lab partner, and then write a group response, therefore requiring both collaboration and metacognition. Several example questions are listed in Table~\ref{refqtable}. The questions were relevant to the experimental work required for that lab, and use similar terminology. For example, in the first lab, students make and then investigate several predictions, which provides scaffolding for the question, ``Why are predictions so useful and important in experimental sciences like physics?'' Students typically took 10-20 minutes to think, discuss, and write responses to these questions. Students were instructed to write a paragraph-long response which, in practice, meant that most responses were three to seven sentences long, approximately. We refer to this third curriculum as RealTime Physics + Reflections.

Five of the first six such reflections are presented in table~\ref{refqtable}. For brevity, surrounding text that unpacked and contextualized the questions has been removed. In pairs, students would write paragraph-long answers to these questions, which would be graded using a rubric. Alongside each reflection question, we also present an E-CLASS item that is closely aligned with the reflection question. During the fifth, and after the sixth week of labs, the reflection questions tackled topics such as the usefulness of simulations, the importance of varying only one variable at a time, and the process of theories becoming accepted in science. Since these topics are not closely aligned with E-CLASS items, they are not presented here.

\subsection{Students}

At our university, we offer two sets of physics courses: algebra-based and calculus-based. Most students in the algebra-based physics courses are health science majors (e.g., pre-medical students), while the calculus-based physics courses mostly enroll engineering and physical science students. These two streams, algebra-based and calculus-based, also have different physics lab courses. Thus, an algebra-based lab is made up mostly of health science majors. However, since engineering students are not required to take the introductory physics lab, the calculus-based lab course is made up mostly of physical science majors. Nonetheless, the lab curriculum is indistinguishable between the two streams, so that the only difference between the algebra-stream and calculus-stream labs is the pool of students enrolled in them. This research was carried out in accordance with the principles outlined in University of Pittsburgh Institutional Review Board (IRB) ethical policy with approval number/ID IRB: PRO15070212.

During the four semesters (Fall 2018 to Spring 2020) that we collected data, we adopted RealTime Physics in a piecewise fashion across the algebra- and calculus-stream labs. For the first two semesters, the algebra-stream labs used the traditional lab curriculum while the calculus-stream labs used the RealTime Physics curriculum. During the third semester, all labs used the RealTime Physics curriculum. During the fourth semester, all labs used the RealTime Physics + Reflections curriculum.

Based on our experience with introductory physics labs at our university and the research literature, we have noticed three characteristics of students who take the labs that might account for variation in how much students learn in the course. First, there are somewhat different pools of students in the fall and spring semesters for this lab course (similar to `on-sequence' and `off-sequence' enrolment~\cite{AltersCounseling}). Students in the spring semester are slightly more likely to be pursuing a pre-health science academic track (e.g., pre-medicine), are slightly more likely to be concurrently enrolled in the second semester of the physics lecture course while taking the lab, and have marginally higher high school grade point average (GPA), SAT Math scores, and grades from the first semester of physics. Second, we note that students who have stronger academic preparation are sometimes able to engage more productively in the lab~\cite{ShaferCategorization}. Third, we note gender differences in how students approach and conduct their lab-work~\cite{DanielssonLabs}. This suggests that, in attempting to understand the impact of our different curricula on student attitudes toward experimental physics, it will be important to account for differences in students' gender, academic preparation, and the semester in which they are enrolled.

\subsection{Other Factors}

During the final semester in which we collected data, our university switched to emergency remote instruction due to the Covid-19 pandemic. As a result, the last five weeks of lab were completed asynchronously and individually, using simulations, rather than in pairs in-person. The pre-lab and homework were completed as normal, and 97\% of students were able to complete all the lab-work remotely. Remote instruction at our university began March 23, and the E-CLASS survey was completed by students April 2-10. In the survey, we asked students about their attitudes toward the physics experiments they had conducted in the lab. Therefore, since students were exposed to a relatively short window of simulation-based asynchronous labs, and were asked to respond to the E-CLASS survey based on their in-lab experiences (which they had mostly completed), we do not believe that student responses during this fourth semester are substantially different because of the switch to emergency remote instruction. For example, we found no significant difference in post FCI scores (see below) for students who complete the semester remotely ($17.4 \pm 0.4$) compared with those from previously semesters ($17.4 \pm 0.4$).

The introductory physics labs at our university are run by graduate student teaching assistants (TAs). The TAs take a course on physics pedagogy during their first year in graduate school~\cite{MarshmanTAGrading}, and receive professional development related to their work supporting student learning in the labs during weekly lab TA meetings. This professional development focused on helping TAs support inquiry learning and inclusion in the labs, and is described in plenty of detail elsewhere~\cite{DoucettePDforLabTAs}. During the four semesters that data was collected, the professional development offered to lab TAs evolved only slightly. For example, during the first semester the TAs wrote reflections during one lab TA meeting, whereas in subsequent semesters they discussed their reflections verbally. Thus, we do not believe that slight evolutions in the lab TA professional development would have impacted student survey responses.

Moreover, the second author provided careful supervision of TAs, including leading lab TA meetings, evaluating performance, and providing feedback throughout the semester. This supervision worked to ensure that students in all lab sections had a consistent, intentional educational experience.

\subsection{Model and Mathematics}

We propose a linear model~\cite{GelmanStatsRegression,TheobaldGeneralizedLinearModels} to predict students' end-of-semester ($PostScore$) E-CLASS scores based on which version of the curriculum they experienced ($Inquiry$, $Reflection$). To account for the possibility of other significant factors impacting our results, and to minimize the potential impact of omitted variable bias~\cite{WalshOmittedVariableBias}, we also include variables for students' start-of-semester E-CLASS scores, the course (either algebra-stream or calculus-stream), the semester (either fall or spring), the student's high school GPA, and the student's gender. The model predicts coefficients ($\beta$), which may be interpreted as the predicted change in $PostScore$ associated with the factors, with all other factors held constant. Mathematically, this model takes the following form:

\begin{equation} \label{modeleq}
\begin{aligned}
    PostScore_i = \; & \; \beta_0 + \beta_1 \, Inquiry_i + \beta_2 \, Reflection_i \\ 
    & + \beta_3 \, PreScore_i + \beta_4 \, Course_i + \beta_5 \, Semester_i \\
    & + \beta_6 \, GPA_i + \beta_7 \, Gender_i + \epsilon_i
\end{aligned}
\end{equation}

$PostScore$ and $PreScore$ are student scores from the E-CLASS, a validated ``expectations and epistemology''~\cite{ZwicklECLASSDevelopment,WilcoxECLASSValidation} survey that aims to measure the extent to which students hold expert-like attitudes toward experimental physics. A total of 30 items are answered on a 5-point Likert scale, with +1 for each expert-like response (e.g., responding ``strongly agree'' or ``agree'' to an item for which the expert response is agreement), 0 for each neutral response, and -1 for each novice-like response (e.g., responding ``strongly disagree'' or ``disagree'' to an item for which the expert response is agreement~\cite{WilcoxECLASSValidation}. The theoretical range of scores on E-CLASS is from -30 to +30. The validation of E-CLASS found an average score of 15.8 on the pre and 14.4 on the post for a national (USA) sample~\cite{WilcoxECLASSValidation}. In interpreting the models in the results section, the coefficient of $PreScore$ indicates the correlation between $PreScore$ and $PostScore$. For example, a coefficient of 0.73 means that each point a student scores on the pre predicts 0.73 points on the post, on average, with all other variables held constant.

$Inquiry$ is an indicator variable, taking on a value of 0 if the student was enrolled in a lab with the traditional curriculum and 1 if the student was enrolled in a lab that used the RealTime Physics curriculum (either with or without the additional reflection questions). $Reflection$ is an indicator variable that takes a value of 1 if the student was enrolled in a lab that used the RealTime Physics + Reflections curriculum, and 0 otherwise. For both these variables, the coefficients indicate the number of additional points students earn on $PostScore$ because they are enrolled in a lab that uses this curriculum, all else held constant.

$Course$ is an indicator variable that takes a value of 0 for algebra-stream students and 1 for calculus-stream students. $Semester$ is an indicator variable that takes a value of 0 for students enrolled in the fall semester and 1 for students enrolled in the spring semester. For these variables also, the coefficients indicate the number of additional points students are predicted to get on the post score because they are enrolled in a calculus-stream or spring semester class, all else held constant.

$GPA$ is the student's high school grade point average, retrieved from university records through an IRB-approved process that anonymizes student data to preserve privacy. The intent of this variable is to provide an approximate accounting for the student's academic preparation. As a variable, $GPA$ has been standardized to have a mean of 0 and a standard deviation of 1. Therefore, the coefficient of GPA indicates the number of additional points a student scores on $PostScore$ corresponding to an increase in $GPA$ by 1 standard deviation. 

$Gender$ is a variable that represents the student's gender, with women assigned 0, men assigned 1, and other gender identities excluded from this analysis because no meaningful statistics could be made for students with non-binary gender identities. The coefficient indicates the number of additional points on $PostScore$ for men, on average, with everything else held constant. We acknowledge that gender is fluid and non-binary, but note that since gender is an important factor that can affect students' experiences in the lab, we should seek to account for it as best as we are able. In this case, we retrieved gender data from university records where it was stored as ``male'', ``female'', or ``other/unknown''. We use the terms ``women'' and ``men'', rather than the terms used in our university record system, to emphasize that we are talking about whole humans rather than biological categories.

As a secondary analysis, we consider the impact of the three curricula on only the five E-CLASS items from Table~\ref{refqtable} that are closely aligned with the reflection questions. We use Eq.~\ref{modeleq}, but note that the $PreScore$ and $PostScore$ in this analysis will have a theoretical range from -5 to +5.

In addition to the above analyses, we also look at the impact of the above factors on students' conceptual learning. For this, we use the Force Concept Inventory (FCI)~\cite{HestenesFCI}. While the only other study of RealTime Physics cited above uses the FMCE~\cite{RealTimePhysics}, most of the other work on inquiry-based introductory physics labs has used the FCI. Furthermore, our department has used the FCI in lecture courses for many years, and have found typically found increases of 5-6 points as students complete the first semester of physics. The FCI consists of 30 items, each of which is either correct or incorrect, resulting in a score between 0 and +30. To evaluate student conceptual learning, we use Eq.~\ref{modeleq}, but with the FCI scores used as $PreScore$ and $PostScore$ intead of E-CLASS scores. Notably, since students take our lab course only after they have completed their first semester of physics lecture (i.e., mechanics), no students in this analysis are concurrently taking a physics lecture course that is teaching mechanics concepts. Thus, unlike other studies that have been unable to distinguish between the impact of lab and lecture on student learning of physics concepts, this analysis suggests the impact of the physics lab on student learning of the force concept, independent of lecture.

We calculated linear models using R Studio. All models were checked to ensure that errors were normally distributed and heteroscedastic. Each model was checked to evaluate the possibility of statistically significant interaction terms. No statistically significant interactions were found in any of the models.

Parsimonious models were developed by iteratively removing non-significant ($p<0.05$) factors until all remaining factors are significant, with similar results. We present both complete and parsimonious models in this manuscript. However, we acknowledge the trade-off between fit parameters and parsimony when seeking to maximize explanatory power in model selection. A 'kitchen sink' model accounts for much but explains little, while an overly simple model may miss important covariates. One approach that is being used increasingly in physics education research (e.g.,~\cite{NissenAIC}) is to compare models using the Akaike Information Criterion (AIC)~\cite{GelmanStatsRegression}, which estimates prediction error while penalizing model complexity, and to report those models that minimize the AIC. In all the models presented in this paper, the complete model minimized the AIC, and is therefore the model that we use to report results. However, for completeness, we report results from both the complete model as well as the parsimonious model. We adopt a criterion of p < 0.05 for statistical significance in this paper, and report uncertainty ranges as 95\% confidence intervals.

\section{Results}

We collected data over four semesters, from fall 2018 to spring 2020. Students filled out surveys during the lab sessions, and were given a small grade incentive for doing so. Students completed the E-CLASS pre on the first day of lab and the post on the second-to-last day. The FCI pre was also done on the first day of lab, and the post was done in week 7, after the mechanics labs were finished. 

Table~\ref{paneldata} provides panel data about survey respondents according to their course, semester, and curriculum. We have fewer responses from the calculus-stream labs because these labs have lower enrolment. Engineering students, who make up the majority of students in our calculus-stream physics lecture sequence, do not take the physics lab at our university). 

Table~\ref{eclassdatabreakdown} indicates the number of pre and post E-CLASS surveys that were collected for each course type, semester, and curriculum. The attrition from pre to post is largely due to students dropping out of the lab course during the first couple weeks, and to students skipping lab during the final weeks of the semester. The number of responses that could be included in the analysis was further decreased because students did not correctly indicate their student ID number, which prevented us from matching pre and post responses. Additionally, at this stage, survey responses that were incomplete or that had patterns suggesting the student did not provide honest answers (e.g., if all "A"s were selected) were removed. There was no statistical difference between average E-CLASS scores from survey responses that were included (pre: $15.4 \pm 0.6$, post: $13.6 \pm 0.6$) and those that were excluded (pre: $15.0 \pm 0.8$, post: $13.2 \pm 1.0$). FCI survey response rates were comparable. Only matched data is presented in this analysis of E-CLASS data from 701 students and FCI data from 569 students.

Table~\ref{eclassmodel} shows results from out linear model for post E-CLASS score. Aside from the pre E-CLASS score, we find two statistically significant factors that predict post E-CLASS scores. $Inquiry$ is not a statistically significant predictor, indicating that students' E-CLASS scores are not different whether the traditional or RealTime Physics curriculum was used. However, $Reflection$ is a statistically significant predictor of post E-CLASS score. Thus, adding reflection questions to the RealTime Physics curriculum is associated with an increase of $3.80 \pm 2.14$ points on the E-CLASS post score, all else being equal. 

In the the models on Table~\ref{eclassmodel}, we find that students' gender is a statistically significant predictor of post E-CLASS score. Men scored $1.46 \pm 1.04$ points higher on the post E-CLASS than women, independent of their pre E-CLASS, high school GPA, and other factors related to the nature of their lab. The final statistically significant factor in Table~\ref{eclassmodel} is $PreScore$, which significantly predicts $PostScore$.

\begin{figure*}
\includegraphics[width=\textwidth]{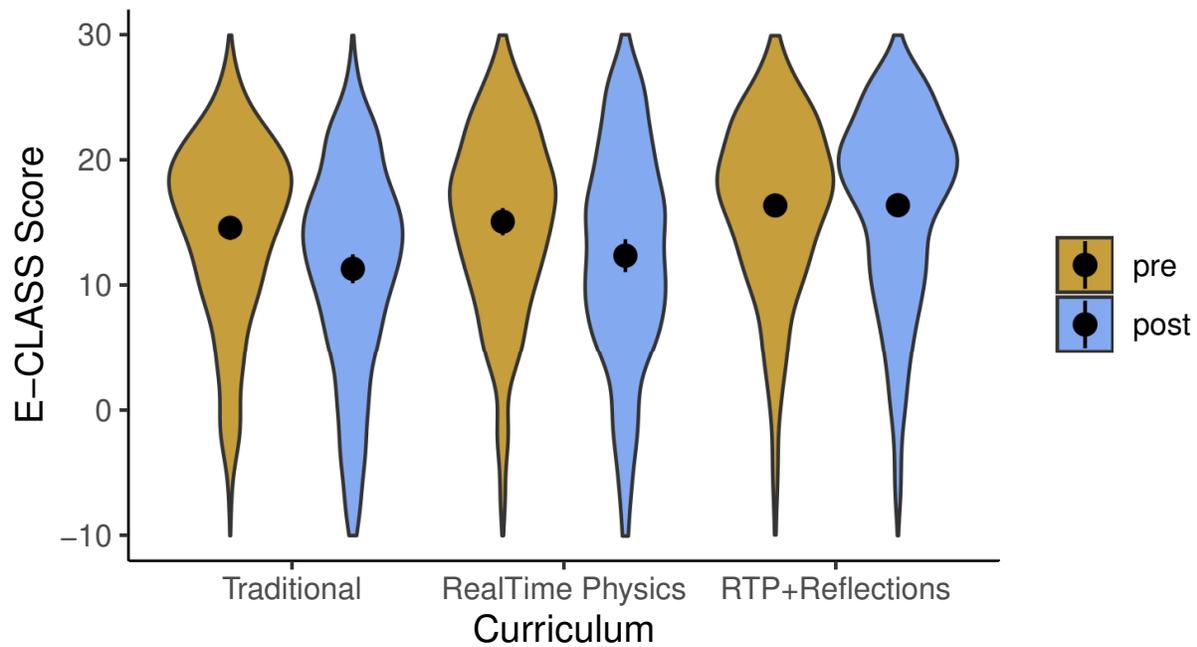}
\caption{Violin plot of matched pre/post E-CLASS scores by semester. The width of the `violin' indicates the frequency of responses with this E-CLASS score, with smoothing (bandwidth = 2). Error bars indicate 95\% confidence intervals for the mean values. E-CLASS scores decrease in each of the first three semesters, but are stable in the RealTime Physics + Reflections semester.\label{eclassviolins}}
\end{figure*}

We may also compare mean scores (not model results) in order to interpret these results. We find that students in labs that used the either the traditional or RealTime Physics curricula (without reflection questions) averaged a pre E-CLASS score of $14.3 \pm 0.7$ and a post E-CLASS score of $11.4 \pm 0.9$ (out of a maximum score of 30). Meanwhile, students in labs that used the RealTime Physics + Reflections curriculum averaged pre E-CLASS scores of $16.4 \pm 0.8$ and post E-CLASS scores of $16.3 \pm 0.9$ as well. To visualize these results, E-CLASS scores from students are presented according to the three curricula in Fig.~\ref{eclassviolins}.

In a second analysis, based on results presented in Table~\ref{refmodel}, we find that the RealTime Physics + Reflections curriculum predicts a total advantage of $1.47 \pm 0.85$ points compared with the Traditional curriculum (i.e., 0.33 + 1.04) for students' scores on a 5-item subset of E-CLASS items related to the reflection questions. As with overall E-CLASS scores, the difference comes from an averted decrease rather than an increase for the RealTime Physics + Reflections curriculum, compared to the traditional curriculum. Mean scores on the 5-item subset decreased from a pre of $1.20 \pm 0.21$ to a post of $0.12 \pm 0.25$ with the Traditional curriculum, but were unchanged, remaining at $1.68 \pm 0.24$, with the RealTime Physics + Reflections curriculum.

In order to understand how the three different curricula may have impacted students' E-CLASS scores, Fig.~\ref{eclassitems} shows the average response for each of the 30 items on the E-CLASS instrument, pre and post, for each of the three curricula. Overall, the average scores decreased from pre and post for most items, for students in labs that used the traditional and RealTime Physics curricula. However, average scores for students in labs that used the RealTime Physics + Reflections curriculum did not seem to decrease as much. Notably, the items related to the reflection questions (23, 3, 5, 14, 16) are not primarily responsible for the difference in E-CLASS post scores for students in labs that used the RealTime Physics + Reflections compared with students in labs that used the RealTime Physics curriculum.

In a follow-up to the main research question, we performed a similar analysis on students' post FCI scores. These model results are presented in Table~\ref{conceptmodel}. Once controlled for course, semester, preparation, and gender, we find that students in labs that used the RealTime Physics inquiry-based curriculum (with or without reflection questions) were predicted to have a post FCI score that was $2.60 \pm 2.58$ points higher. In terms of mean FCI scores, this reflects an increase from $16.7 \pm 0.3$ to $17.8 \pm 0.3$, on average, for students enrolled in labs that used the RealTime Physics conceptual inquiry-based curriculum, compared with a decrease from $15.5 \pm 0.7$ to $14.8 \pm 0.7$, on average, for students enrolled in labs that used the traditional curriculum (out of a maximum score of 30). 

Students enrolled in the calculus-stream physics labs had post FCI scores that were $1.89 \pm 1.38$ points higher than students in algebra-stream physics labs, controlled for pre FCI score. This difference reinforces the importance of including potential covariates such as $Course$ in the analysis. The pre scores on FCI significantly predicted post scores. No other factors were statistically significant in our model.

Unlike the complete model (which is preferred because it minimizes the AIC), the parsimonious model for post FCI score indicates a statistically significant decrease to FCI scores for students who enrolled in labs that used the RealTime Physics + Reflections curriculum. This effect is an artifact of removing the covariates for the parsimonious model, and reflects weak correlations between the removed covariates and the Reflection variable. This is another reason that we prefer the complete models than the parsimonious models for reporting results.

\section{Discussion}

In our introductory physics labs, we found that students enrolled in labs that used a conceptual inquiry-based curriculum demonstrated the same decrease in the level of expert-like thinking on an assessment of their attitudes toward experimental physics as students who were enrolled in labs that used a traditional, highly-structured curriculum. However, students who were enrolled in a lab that used the conceptual inquiry-based lab curriculum supplemented by additional reflection question avoided such a decrease. A similar pattern of non-decreasing E-CLASS scores has been reported by other researchers for ``open-ended'' labs, labs that ``develop lab skills'', and integrated approaches to lab curricula such as ISLE, SCALE-UP, or Studio Physics~\cite{WilcoxOpenEnded,WilcoxDevelopingSkills,WilcoxImpactApproach}. These results suggest that inquiry-based labs that include reflection on the nature of science and ways of knowing in experimental physics may also be suitable for colleges and universities that seek to transform lab instruction while attending to students' beliefs and attitudes related to experimental physics.

\begin{figure*}
\includegraphics[width=\textwidth]{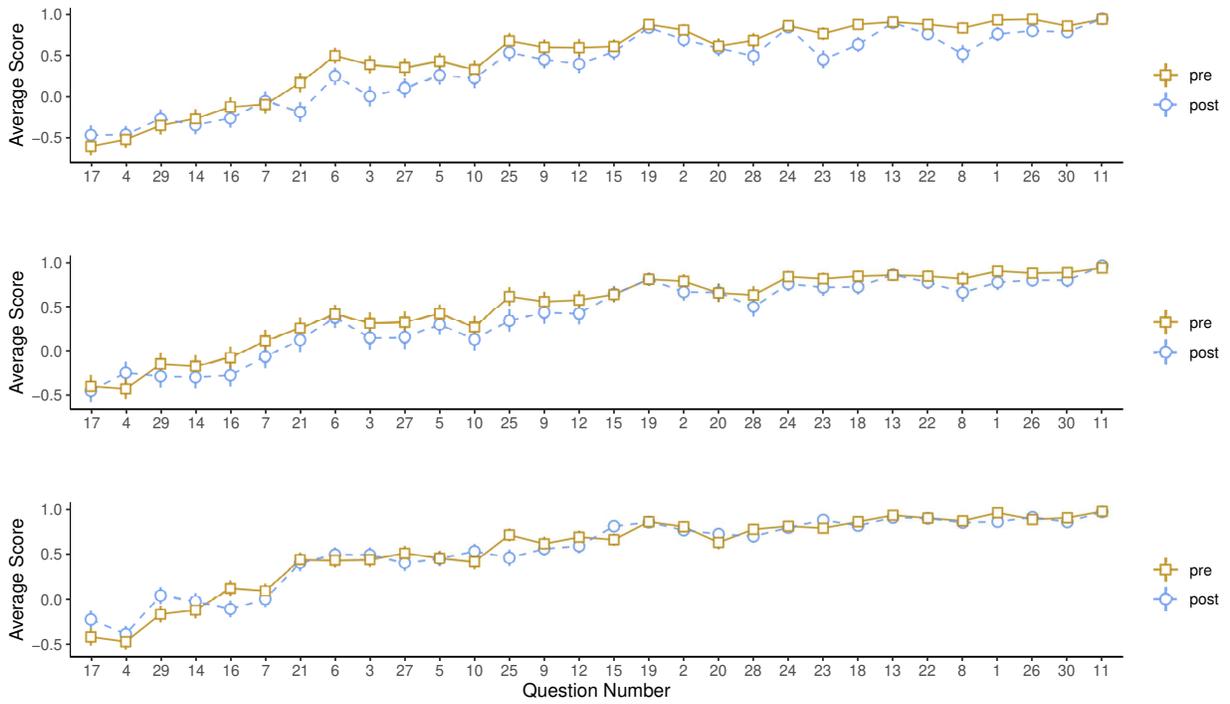}
\caption{Average item-by-item pre/post E-CLASS results for three curricula. The order of the 30 items on the E-CLASS instrument matches \cite{WilcoxECLASSValidation} for comparison. Error bars show 95\% confidence intervals. Comparing student responses from the Traditional (top) and RealTime Physics (middle) curricula indicates a difference in which items are affected, but both curricula show a persistent decrease in expert-like responses across most items. In comparison, responses from students in the RealTime Physics + Reflections (bottom) curriculum show minimal decrease in expert-like responses on the 30 E-CLASS items.\label{eclassitems}}
\end{figure*}

In designing the additional reflection questions for our conceptual inquiry-based labs, we sought to align our students' lab-work with the goals we identified for the course. Five of the first six reflection questions (see Table~\ref{refqtable}) aligned neatly with E-CLASS items. Students who enrolled in the lab with the RealTime Physics + Reflections curriculum scored $1.47 \pm 0.85$ points better on these items than their peers who enrolled in the lab with the Traditional curriculum. However, since the overall E-CLASS benefit from the RealTime Physics + Reflections curriculum was $3.80 \pm 2.14$ points, compared with the Traditional curriculum, the results suggest that the effect of the novel curriculum was somewhat more than students simply responding to narrowly framed prompts. Furthermore, an item-by-item analysis of results (see Fig.~\ref{eclassitems}) shows that the difference between E-CLASS scores for students enrolled in labs that used the RealTime Physics curriculum and students enrolled in labs that used the RealTime Physics + Reflections curriculum is more diffuse than just those E-CLASS items aligned with the reflection questions. This suggests that the novel curriculum may have had an impact on how students thought about experimental physics. The process of revising lab-work to align with course goals may worthwhile for other instructors who seek to improve or transform their lab courses. 

We also found a statistically significant difference between the post E-CLASS scores of men and women, when controlling for pre E-CLASS scores. This result aligns with previous findings from a large-scale study of E-CLASS scores~\cite{WilcoxECLASSgender}, which found that women in first year physics labs who were not physics majors had lower post E-CLASS scores than men, controlling for pre E-CLASS score. Qualitative research on this topic suggests that gendered interactions between students within the masculinized culture of physics may be responsible for students having different learning experiences depending on their gender~\cite{DoucetteHermione,Doucette4WomenEJP,DoucetteGoodLabPartner,MarshmanLongitudinal,HolmesGenderLabs,DayGenderLabs,DanielssonLabs,MasculinitiesGonsalves,QuinnGenderLabs,QuinnGenderRoles}.

Past investigations of the impact of inquiry-based lab instruction using the FCI have either reported no significant difference between traditional and inquiry-based instruction~\cite{NockInquiryLab, RiazDLSCL} or studied labs in which students were simultaneously enrolled in a physics lecture course, making it difficult to determine the impact of the lab by itself~\cite{RealTimePhysics,VanDomelenLabFCI}. Our results suggest that students enrolled in a RealTime Physics conceptual inquiry-based lab had post FCI scores that were $2.60 \pm 2.58$ points higher than students enrolled in traditional labs. This increase is comparable to (and is in addition to) the 5-6 point increase that students typically achieve in introductory physics lecture courses (there is some variation between instructors, course, semester, etc.). However, lab is worth fewer credits, and students typically invest less time in the lab than in their physics lecture courses, so we estimate that FCI-gain-per-credit or FCI-gain-per-hour-invested is comparable between lab and lecture for students enrolled in labs that used the RealTime Physics curriculum.

We note that the E-CLASS scores of students enrolled in a traditional lab class decrease during the course of the semester. For the E-CLASS, it may be possible to attribute this retrenchment to students coming to see that lab-work is less authentic than they had originally imagined, or less authentic than other labs (e.g., biology or chemistry) they have taken in college. For example, students may come to agree with the item, ``The primary purpose of doing a physics experiment is to confirm previously known results,''~\cite{ZwicklECLASSDevelopment} from E-CLASS, shifting to novice-like thinking because their experimental work primarily seems focused on confirming theory they have previously learned in class~\cite{Hu2017ECLASS}.

There are several important limitations to this analysis. The E-CLASS provides only a superficial measure of students' epistemological views about experimental physics. It is not clear the extent to which students' views are being deeply or enduringly shifted as a result of the reflection questions and/or the use of a conceptual inquiry-based curriculum. These data also represent only one selective institution, with a particular set of policies for running the lab course, which may limit the extent to which these results may be extrapolated to other institutions and introductory lab configurations. Two approaches to lab-work that were not explored are the Labatorial~\cite{AhrensmeirLabatorials,KalmanLabatorial,SobhanzadehLabatorials}, which synergizes lab and recitation in a way that might produce similar results to lab-lecture integrations, and online lab-work~\cite{MircikVirtualLabs}.

Our results suggested that students enrolled in a lab that used a conceptual inquiry-based curriculum may have demonstrated the same decrease in expert-like attitudes toward experimental physics as students enrolled in a lab that uses a traditional physics curriculum. However, when the conceptual inquiry-based curriculum was supplemented with reflection questions that addressed the nature of science and ways of knowing in experimental physics, students' attitudes toward experimental physics remained stable. In addition, our results suggested that students enrolled in a stand-alone lab that used a conceptual inquiry-based curriculum may have demonstrated improved understanding of concepts in mechanics compared with students enrolled in a stand-alone lab that used a traditional, highly-structured curriculum.

Instructors who seek to transform their introductory physics labs have a wealth of curricular approaches available. Integrated lecture and lab approaches such as ISLE, SCALE-UP, or Studio Physics have been shown to improve student understanding of physics concepts (e.g.,~\cite{EtkinaDesignReflection,DoriTEAL}) as well as avoid a decrease in their attitudes toward experimental physics~\cite{WilcoxImpactApproach}. ``Open-ended''~\cite{WilcoxOpenEnded} labs and labs that ``develop lab skills''~\cite{WilcoxDevelopingSkills} are likewise effective at keeping students' attitudes toward experimental physics stable. Our results suggest that a conceptual inquiry-based lab curriculum that includes reflection questions may likewise return stable E-CLASS results.

\begin{acknowledgments}
We wish to thank the faculty who helped identify goals for the lab, the graduate student teaching assistants who helped to administer surveys, the administrative workers in our department who supported our work, and Bob Devaty for insightful comments about this manuscript.

\end{acknowledgments}

\bibliography{the}

\newpage

\begin{table}[htbp]
\caption{The 12 labs included in the RealTime Physics and RealTime Physics + Reflection curricula, from~\cite{RealTimeBook1,RealTimeBook2,RealTimeBook3,RealTimeBook4}. \label{labstable}}
\begin{ruledtabular}
\begin{tabular}{cl}
Week & Lab\\
 \hline
1 & Changing Motion\\
2 & Force and Motion\\
3 & Combining Forces\\
4 & Newton's Third Law and Conservation of Momentum\\
5 & Two-Dimensional Motion\\
6 & Conservation of Energy\\
7 & DC Circuits\\
8 & Capacitors and RC Circuits\\
9 & Magnetism and Electromagnetism\\
10 & Reflection and Refraction\\
11 & Geometrical Optics\\
12 & Waves of Light\\
\end{tabular}
\end{ruledtabular}
\end{table}

\begin{table}[htbp]
\caption{Reflection questions from five of the first six labs. Students are asked to write a paragraph response to one such item in each week's lab report. For sake of brevity, surrounding text that unpacked and contextualized the questions has been removed. In addition, E-CLASS items~\cite{ZwicklECLASSDevelopment} that align with the questions are indicated. These items are analyzed later.\label{refqtable}}
\begin{ruledtabular}
\begin{tabular}{p{0.1\linewidth}p{0.4\linewidth}p{0.4\linewidth}}
Week & Question & E-CLASS item\\
 \hline
1 & Why are predictions so useful and important in experimental sciences like physics? & When I am doing an experiment, I try to make predictions to see if my results are reasonable. \\
2 & Why is it important to think about sources of systematic error when doing physics experiments? & When doing a physics experiment, I don't think much about sources of systematic error. \\
3 & Why is it important for scientists to use uncertainties when they analyze and share their work? & Calculating uncertainties usually helps me understand my results better. \\
4 & Why is it important that physics students have opportunities to come up with their own experiments to investigate? & When doing an experiment I usually think up my own questions to investigate. \\
6 & Some people believe that the purpose of doing a physics lab is simply to verify facts about physics that you already know... What is the goal of the physics lab? & The primary purpose of doing a physics experiment is to confirm previously known results. \\
\end{tabular}
\end{ruledtabular}
\end{table}

\begin{table}[htbp]
\caption{Number of responses for our two survey instruments, broken down according to course type (algebra or calculus) and semester (fall or spring) and separately broken down according to three curricula. The core results in this paper rely on E-CLASS scores from three different curricula, which are reasonably well balanced. FCI scores, used for a secondary analysis, are drawn from data that is less well balanced, reducing the statistical power of our results. \label{paneldata}}
\begin{ruledtabular}
\begin{tabular}{rcc}
 & E-CLASS & FCI \\
 \hline
algebra, fall & 254 & 134 \\
algebra, spring & 392 & 348 \\
calculus, fall & 25 & 38 \\
calculus, spring & 30 & 49 \\
 \hline
traditional & 208 & 73 \\
RealTime Physics & 175 & 196 \\
RealTime Physics + Reflections & 318 & 300 \\
 \hline 
total & 701 & 569 \\
\end{tabular}
\end{ruledtabular}
\end{table}

\begin{table}[htbp]
\caption{Number of responses to the E-CLASS survey to illustrate response rate and gender composition. The first two columns indicate the number of survey responses collected. The third column indicates the responses that had both pre and post survey complete, and could be matched (i.e., the student ID was correct on both). The fourth column indicates the fraction of students that were men for each course type, among the surveys that were included in this analysis.\label{eclassdatabreakdown}}
\begin{ruledtabular}
\begin{tabular}{rcccc}
 & Pre & Post & Included & Men \\
 \hline
algebra, fall & 371 & 312 & 254 & 0.37 \\
algebra, spring & 603 & 460 & 392 & 0.35 \\
calculus, fall & 71 & 58 & 25 & 0.52 \\
calculus, spring & 50 & 36 & 30 & 0.59 \\
 \hline
traditional & 450 & 237 & 208 & 0.38 \\
RealTime Physics & 275 & 245 & 175 & 0.38 \\
RealTime Physics + Reflections & 360 & 382 & 318 & 0.34\\
 \hline 
total & 1085 & 864 & 701 & 0.38 \\
\end{tabular}
\end{ruledtabular}
\end{table}

\begin{table}[htbp]
\caption{Coefficients from complete (first, preferred) and parsimonious (second) models for E-CLASS PostScore (n = 701), along with standard errors and p-values (see eq.~\ref{modeleq}). Both models predict PostScore based on the curriculum (Inquiry and Reflection) along with control variables described in the text. In these models, the RealTime Physics + Reflections curriculum predicts an increase in E-CLASS PostScore by more than 3 points. The models additionally suggests that gendered effects might be relevant for E-CLASS scores.\label{eclassmodel}}
\begin{ruledtabular}
\begin{tabular}{lcccccc}
Factor & $\beta$ & SE & p & $\beta$ & SE & p\\ 
 \hline
Inquiry & 0.51 & (0.81) & 0.53\\
Reflection & 3.80 & (1.07) & $<$ 0.001 & 3.69 & (0.49) & $<$ 0.001\\
 \hline
intercept & -0.15 & (0.79) & 0.85 & -0.05 & (0.61) & 0.93\\
PreScore & 0.73 & (0.04) & $<$ 0.001 & 0.74 & (0.04) & $<$ 0.001\\
Course & 1.33 & (0.93) & 0.15\\
Semester & -0.58 & (0.89) & 0.52\\
GPA & 0.06 & (0.25) & 0.82\\
Gender & 1.46 & (0.52) & 0.005 & 1.49 & (0.51) & 0.004\\
 \hline
$R^2$ & \multicolumn{3}{c}{0.46} & \multicolumn{3}{c}{0.46} \\  
\end{tabular}
\end{ruledtabular}
\end{table}

\begin{table}[htbp]
\caption{Coefficients from complete (first, preferred) and parsimonious (second) models for E-CLASS Reflection Question PostScore (n = 701), along with standard errors and p-values (see eq.~\ref{modeleq}). The models predict the score from fix items associated with the reflection questions. In these models, the RealTime Physics curriculum predicts an increase in the student's score on 5 E-CLASS items by 0.33 points, and the RealTime Physics + Reflections curriculum predictions an additional 1.04 point increase.\label{refmodel}}
\begin{ruledtabular}
\begin{tabular}{lcccccc}
Factor & $\beta$ & SE & p & $\beta$ & SE & p\\ 
 \hline
Inquiry & 0.33 & (0.24) & 0.17 & 0.41 & (0.20) & 0.04\\
Reflection & 1.04 & (0.35) & 0.001 & 0.92 & (0.18) & $<$ 0.001 \\
 \hline
intercept & -0.60 & (0.20) & 0.003 & -0.67 & (0.15) & $<$ 0.001 \\
PreScore & 0.48 & (0.04) & $<$ 0.001 & 0.48 & (0.04) & $<$ 0.001 \\
Course & 0.17 & (0.28) & 0.54\\
Semester & -0.13 & (0.27) & 0.64\\
GPA & 0.06 & (0.07) & 0.43\\
Gender & 0.54 & (0.16) & $<$ 0.001  & 0.53 & (0.15) & $<$ 0.001\\
 \hline
$R^2$ & \multicolumn{3}{c}{0.28} & \multicolumn{3}{c}{0.28} \\  
\end{tabular}
\end{ruledtabular}
\end{table}

\begin{table}[htbp]
\caption{Coefficients from complete (first, preferred) and parsimonious (second) models for FCI PostScore (n = 569) (see eq.~\ref{modeleq}). The model predicts FCI PostScore based on the curriculum (Inquiry and Reflection) along with control variables, as described in the text. The RealTime Physics curriculum predicts an increase in FCI score by 2.6 points. \label{conceptmodel}}
\begin{ruledtabular}
\begin{tabular}{lcccccc}
Factor & $\beta$ & SE & p & $\beta$ & SE & p\\
  \hline
Inquiry & 2.60 & (1.29) & 0.04 & 2.46 & (0.69) & $<$ 0.001\\
Reflection & -1.38 & (1.15) & 0.23 & -1.23 & (0.47) & 0.01\\
 \hline
intercept & 3.25 & (1.32) & 0.008 & 3.23 & (0.75) & $<$ 0.001\\
PreScore & 0.74 & (0.04) & $<$ 0.001 & 0.75 & (0.03) & $<$ 0.001\\
Course & 1.89 & (0.69) & 0.006 & 1.85 & (0.63) & 0.004\\
Semester & 0.12 & (1.14) & 0.91\\
GPA & 0.16 & (0.21) & 0.46\\
Gender & 0.02 & (0.47) & 0.97\\
 \hline
$R^2$ & \multicolumn{3}{c}{0.55} & \multicolumn{3}{c}{0.55}\\  
\end{tabular}
\end{ruledtabular}
\end{table}

\end{document}